\documentclass[aps,preprint,superscriptaddress,showpacs]{revtex4-1} %prl,twocolumn
\usepackage{graphicx}
\usepackage{graphicx,epstopdf,color}
\usepackage{amsfonts}
\usepackage{amsmath,amssymb,mathrsfs}
\usepackage{bm}
\usepackage{pst-grad}

\usepackage{dcolumn}

\def \basp{BaFe$_2$(As$_{1-x}$P$_{x}$)$_2$}

\begin{document}
\title{The strange metal Hall effect connects quantum criticality and superconductivity in an iron-based superconductor}

\author{Ian M. Hayes}
\affiliation{Department of Physics, University of California, Berkeley, California 94720, USA}
\affiliation{Materials Science Division, Lawrence Berkeley National Laboratory, Berkeley, California 94720, USA}

\author{Nikola Maksimovic}
\affiliation{Department of Physics, University of California, Berkeley, California 94720, USA}
\affiliation{Materials Science Division, Lawrence Berkeley National Laboratory, Berkeley, California 94720, USA}

\author{Mun K. Chan}
\affiliation{Los Alamos National Laboratory, Los Alamos, NM 87545, USA}

\author{Gilbert N. Lopez}
\affiliation{Department of Physics, University of California, Berkeley, California 94720, USA}
\affiliation{Materials Science Division, Lawrence Berkeley National Laboratory, Berkeley, California 94720, USA}

\author{B. J. Ramshaw}
\affiliation{Los Alamos National Laboratory, Los Alamos, NM 87545, USA}
\affiliation{Laboratory of Atomic and Solid State Physics, Cornell University, Ithaca, NY, 14853.}

\author{Ross D. McDonald}
\affiliation{Los Alamos National Laboratory, Los Alamos, NM 87545, USA}

\author{James G. Analytis\footnote{For correspondence contact analytis@berkeley.edu}}
\affiliation{Department of Physics, University of California, Berkeley, California 94720, USA}
\affiliation{Materials Science Division, Lawrence Berkeley National Laboratory, Berkeley, California 94720, USA}

%\pacs{74.25.Dw, 74.25.fc, 74.40.Kb}

\begin{abstract}
Many unconventional superconductors exhibit a common set of anomalous charge transport properties that characterize them as `strange metals', which provides hope that there is single theory that describes them.
However, model-independent connections between the strange metal and superconductivity have remained elusive. In this letter, we show that the Hall effect of the unconventional superconductor \basp\ contains an anomalous contribution arising from the correlations within the strange metal.
This term has a distinctive dependence on magnetic field, which allows us to track its behavior across the doping-temperature phase diagram, even under the superconducting dome. These measurements demonstrate that the strange metal Hall component emanates from a quantum critical point and, in the zero temperature limit, decays in proportion to the superconducting critical temperature. This creates a clear and novel connection between quantum criticality and superconductivity, and suggests that similar connections exist in other strange metal superconductors.
\end{abstract}

\maketitle

In a Fermi liquid with only one relaxation rate, the Hall coefficient, $R_H$, should be nearly independent of temperature.\cite{pippard_magnetoresistance_2009} The need to explain the strong temperature dependence that is observed in the Hall coefficient of many high-temperature superconductors (HTSCs) has been a motivating force behind several of the major theoretical approaches to these materials.\cite{anderson_hidden_2011, varma_hall_2003, coleman_how_1996, hussey_phenomenology_2008} A very similar phenomenology has been observed in other unconventional superconductors, including some heavy fermion superconductors and the more recently discovered iron-based HTSCs, which strongly suggests that this phenomenon is intimately connected to the essential physics of unconventional superconductivity.

In addition to an unusual Hall coefficient, these systems all display $T-$linear resistivity in parts of their phase diagrams. Recent studies have found a magnetic analogue of the $T-$linear resistivity, which motivates a careful study of the field dependence of $R_H$.\cite{hayes_scaling_2016, giraldo-gallo_scale-invariant_2018, sarkar_correlation_2019} In this report we present the magnetic field dependence of the Hall resistivity in the prototypical iron-pnictide superconductor \basp. This material displays the essential strange metal transport properties near the zero temperature end point of anti-ferromagnetic/orthorhombic (AFM/ORT) transition, where its superconducting $T_c$ is also maximal. It can be cleanly doped across the entire superconducting dome, and its upper critical fields are low enough that the normal state can be accessed at low temperatures with modern pulsed field technology, enabling us to measure $R_H$ throughout the entire doping-temperature phase diagram. For this study we have focused on the overdoped side, where there is no phase transition that reconstructs the Fermi surface and complicates the interpretation of the Hall coefficient.\cite{kasahara_evolution_2010} Single crystals of \basp\ were grown by self-flux methods described elsewhere\cite{analytis_enhanced_2010, nakajima_growth_2012}, and the Hall resistivity was measured by a standard four-probe, AC lock-in method. Magnetic fields up to sixty-five tesla were accessed at the NHMFL Pulsed Field Facility.% in Los Alamos, New Mexico.

The isovalent phosphorous-for-arsenic substitution has another great advantage, which is that it keeps the system compensated, that is it keeps $n_e = n_h$, throughout the phase diagram. Multiband materials like the iron-based HTSCs can show very complicated field dependencies in $R_H(H)\equiv \rho_{xy}(H)/H$. However, the behavior of compensated systems is much more tightly constrained.
The low-field Hall coefficient should be small in these systems, because the conductivity contributions of holes and electrons nearly cancel.\cite{pippard_magnetoresistance_2009} As the magnetic field grows, $R_H$ should increase in magnitude as it approaches the high field limit, where the Hall coefficient is always determined by the net carriers in the system: $R_H(H \rightarrow \infty) = 1/(en_{total}) = 1/e(n_e-n_h) \approx \infty$. 
Therefore, band theory provides an unusually straightforward prediction for the field dependence of $R_H$ in \basp, specifically that in general it should increase with increasing field. We give a more detailed discussion of this in the supplementary material (SI), and although the  models we use are not exhaustive, the conclusions we make in this paper are model independent.\cite{analytis_enhanced_2010, shishido_evolution_2010} 

\begin{figure*}[ht]
\includegraphics[width=17.8cm]{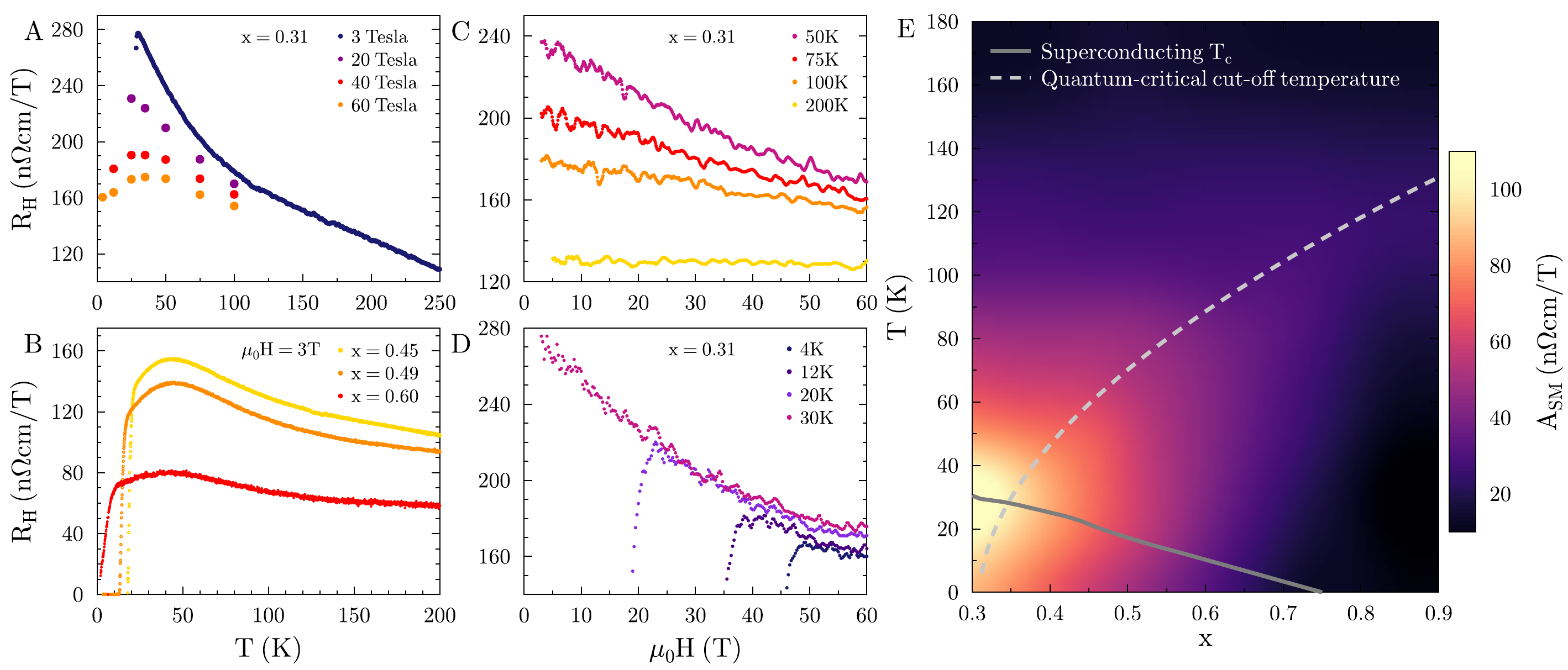} 
\caption{{\bf Strange metal behavior in \basp.} {\bf A.} The magnitude of the Hall coefficient (it's sign is everywhere electron-like) as a function of temperature at optimal doping, evaluated at several fields. The low temperature upturn is suppressed with increasing field  {\bf B.} $R_H$ (evaluated at 3 tesla) versus temperature at several dopings. The upturn feature persists to higher doping, but is suppressed below a cut-off temperature that grows as the doping is increased. {\bf C.} $R_H$ as a function of field at optimal doping. The decrease in $R_H$ is contrary to what is expected from the band structure, and only happens below about 150K. {\bf D.} Below $T_c$, $R_H$ exhibits nearly the same field dependence, indicating a similar size of the strange metal term down to low temperatures. {\bf E} The amplitude of the strange metal term in $R_H$ as a function of $x$ and $T$, estimated from the field dependence of $R_H$ (see main text). The superconducting transition temperature is given by the grey solid line and the white dotted line is a guide for the eye, emphasizing the boundary of the region where where the strange metal $R_H$ is doping independent.}
\label{fig:color} 
\end{figure*}

Figure \ref{fig:color} shows $R_H$ (current in the ab-plane, field along the c-axis) in optimally doped \basp\ as a function of temperature and field. As shown in Figure \ref{fig:color}A, below about 150 kelvin, there is a pronounced enhancement of $R_H$ similar to the $T-$dependence seen in many other strange metals.\cite{hussey_phenomenology_2008} As the system is doped away from optimal doping, this temperature enhancement slowly diminishes, which immediately suggests an association with the proximity to the critical point (Figure \ref{fig:color}B). Within this temperature window, $R_H$ also has an unusual field dependence which is contrary to the expectations of band theory described above; $R_H$ decreases with increasing field (or equivalently enhanced with decreasing field, Figure 1D) in close analogy to the low-temperature enhancement. The parallel between the field and temperature dependence of $R_H$ is reminiscent of the parallel between the $T-$linear and $H-$linear resistivity of this material. \cite{hayes_scaling_2016} 

\begin{figure*}[ht]
\includegraphics[width=17.75cm]{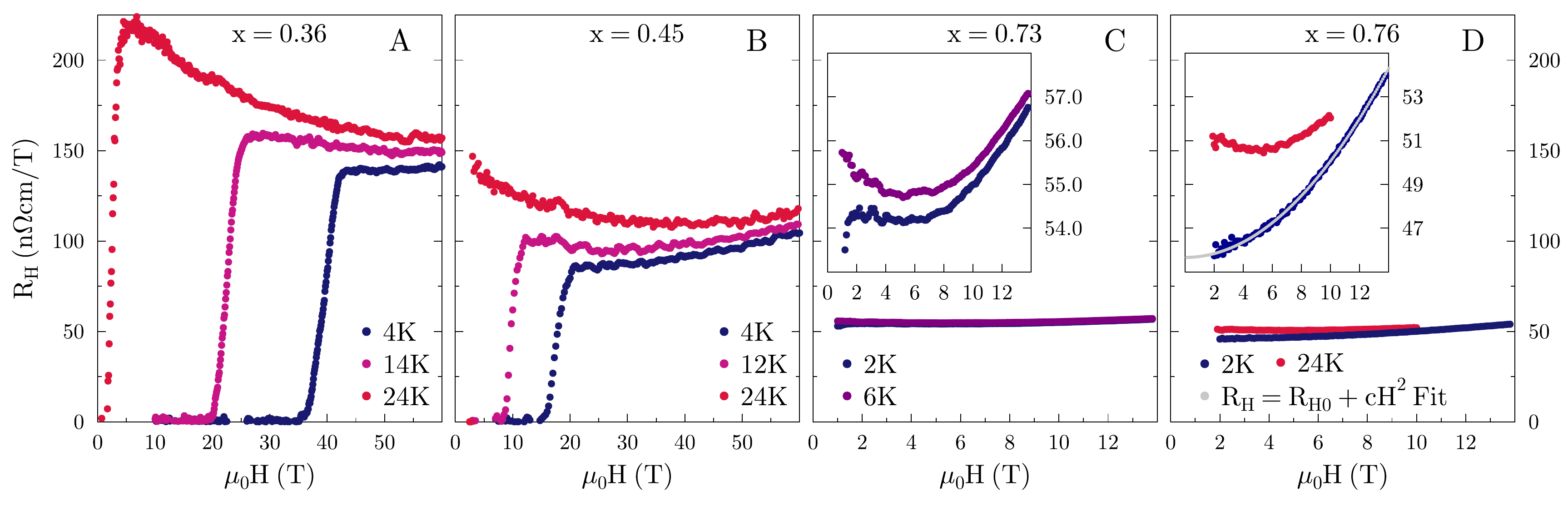} 
\caption{{\bf Low temperature Hall coefficient of \basp.} The strange metal enhancement in $R_H$ is strongest near optimal doping. At moderate doping and high fields the $\sim H^2$ trend due to multiband physics is visible ({\bf B}). At the edge of the superconducting dome ({\bf C}) the strange metal term is still visible, while the first nonsuperconducting sample ({\bf D}) shows pure multiband behavior at low enough temperatures.}
\label{fig:RxyvB} 
\end{figure*}

The connection of the unusual field and temperature dependence of the $R_H$ to the putative quantum critical point can be seen explicitly by studying the evolution of $R_H(T,B)$ as a function of doping $x$. In Figure \ref{fig:RxyvB}A-D we illustrate the evolution of the Hall coefficient for four characteristic compositions $x$. The low field enhancement is dramatically suppressed with doping, similar to suppression of the low temperature enhancement of $R_H$ away from the critical point (Figure \ref{fig:color}B). Moreover, even though the very low field data is cut off by superconductivity, it can be readily seen that the enhancement is suppressed with decreasing temperature. The enhancement only vanishes in samples that are so highly doped that no superconductivity is observed at all.

\begin{figure*}[ht]
\includegraphics[width=17.75cm]{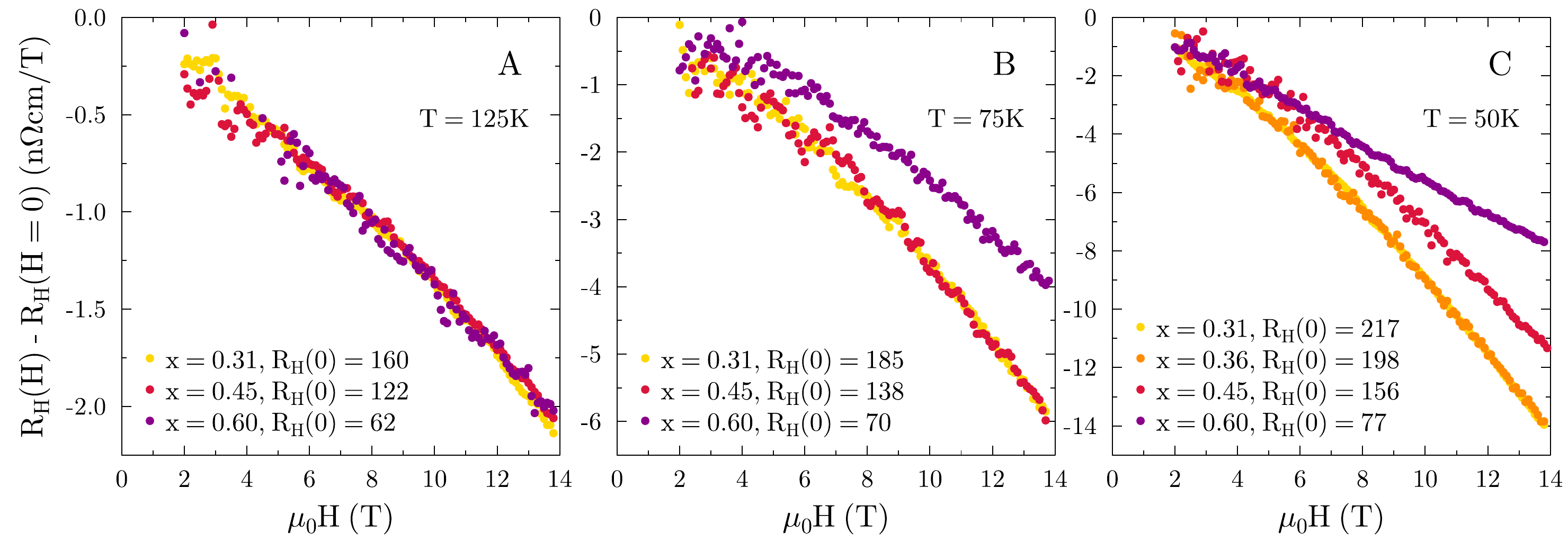} 
\caption{{\bf Low field enhancement of the Hall number across the critical fan.} {\bf A.} The field dependence of $R_H \equiv \rho_{xy}(H)/H$ at 125K for several dopings. The reduction is perfectly uniform in field, indicating that the strange metal part of $R_H$ the same for these samples at this temperature {\bf B.} At 75K the anomalous response weakens for $x = 0.60$, but is still uniform at lower dopings. {\bf C.} By 50K, the window of uniform field dependence is narrower still, with only the $x = 0.31$ and $x = 0.36$ samples matching. Zero field $R_H$ values are given in $n\Omega cm/T$}
\label{fig:fan} 
\end{figure*}

The high temperature field dependence of $R_H$ is shown in Figure \ref{fig:fan}. The zero-field value $R_H(T,0)$ has been subtracted so as to highlight how the low field enhancement of $R_H$ changes between compositions $x$. Strikingly, the low field enhancement is doping independent for a substantial range of dopings, and this range decreases as the temperature is decreased. For example, at 125K (Figure \ref{fig:fan}A), compositions in the range $0.31<x<0.6$ have identical enhancements, while at 50K (Figure \ref{fig:fan}C) only compositions in the range $0.31<x<0.36$ have the same enhancement. The low-field enhancement of $R_H$ evolves as if it was part of a critical fan; within the fan the enhancement is constant as a function of doping, but always decays outside. The behavior of the low-field upturn as a function of doping and temperature (Figure \ref{fig:color}E) not only suggests that it has a quantum-critical origin, but reveals two striking patterns that form the main findings of this work.   

The first of these is that the presence or absence of the low-field enhancement in $R_H$ at the lowest temperatures perfectly correlates with the presence or absence of superconductivity. As shown in Figure \ref{fig:RxyvB}, the field dependence of $R_H$ is strongest near optimal doping. It becomes weaker as one moves towards the overdoped side of the phase diagram, but there is still a discernible decrease with increasing field in all of the superconducting samples we have measured. For samples beyond the superconducting dome, one can always reach a temperature low enough that $R_H$ only increases with field as expected from multiband physics, indicating a complete absence of the strange metal phenomenon (see Figure 2D). Thus, the presence and size of the low field enhancement of the Hall coefficient in the $T=0$K ground state is correlated to the presence of superconductivity and the size of $T_c$.

Our second finding is that there is a fan-like region, emanating from the zero temperature endpoint of the AFM/ORT phase transition, in which the low-field enhancement is a function of temperature but not of doping. This can be readily seen in Figure \ref{fig:fan} shows $R_H(H)$ for three dopings at three elevated temperatures. 
At high temperatures, the field dependence of $R_H$ is the same for all three samples. However, as the temperature is lowered, the doping range over which the field dependence is the same shrinks, with the samples at higher dopings showing a weaker field dependence than those at lower dopings, and eventually vanishes near zero temperature and optimal doping. This pattern is expected of phenomena that have their origin in the fluctuations of the AFM/ORT phase; as long as the temperature is larger than the energy scale which defines the ``detuning" from a stable phase, fluctuations should exhibit a nearly identical spectrum. In this region, the observable consequences of the proximate critical point should be the same. At temperatures below the detuning energy, fluctuations and their effects are suppressed. The absence of any internal detuning scale at the phase transition is what makes the transition critical.\cite{sondhi_continuous_1997}

It is significant that while the low-field enhancement of $R_H$ is invariant within the fan, the absolute value of $R_H$ at zero-field changes considerably across the phase diagram (see Figure \ref{fig:fan}). This makes it unlikely that the strange metal physics behind the enhancement enters as a prefactor to the multiband $R_H$. It is much more likely that the Hall coefficient in \basp\ is composed of a sum of terms, one of which is the result of simple multiband physics and the other arising from strong correlations - the strange metal correction. 
In order to quantify the evolution of this anomalous field dependence in the $T-x$ phase diagram, we decompose $R_H$ into two terms, 
%\begin{equation}
$R_H(H)  = {R_H}^{BS}(H) + {R_H}^{SM}(H)$. Such a decomposition is in keeping with treatments of anomalous contributions to the Hall effect, interpreted as the addition of Hall currents~\cite{nagaosa_anomalous_2010}. Each term
%\label{eq:model}
%\end{equation}
is characterized by its field dependence: decreasing in field for $R_H^{SM}$ and increasing in field for $R_H^{BS}$.
This decomposition is purely phenomenological, and meant only to help identify how this field dependence evolves as a function of doping and temperature. Multiband theory suggests that the field dependence of the band structure term should be $\sim H^2$, and we adopt this form for  $R_H^{BS}$. The field dependence of $R_H^{SM}$ that characterizes the low field enhancement of $R_H$, is described by some decreasing function of field, and we find a decaying exponential $\sim A_{SM}e^{-H/H_0}$ captures this dependence over a wide range of doping and temperature.
Further details of this decomposition are given in the SI, but we emphasize that none of our conclusions hinge on the specific choice of parameterization for these terms because the important trends in $R_H^{SM}(x, T)$ are clearly visible in the raw data (see Figures \ref{fig:RxyvB} and \ref{fig:fan}), as described above. 
 The parameter $A_{SM}$ simply provides a vehicle to quantify the magnitude of the low-field enhancement in $R_H$.  Plotting $A_{SM}$ as a function of doping and temperature (Figure \ref{fig:color}E) neatly communicates the two main findings of this work; the low-field enhancement is invariant within the critical fan, decaying outside it in harmony with the decline of $T_c$. $R_H^{SM}$ can be interpreted as the strange metal correction to the Hall number and at $T=0$K, measures the proximity to the putative quantum critical point.

The fan pattern is exceptionally clear in Figure \ref{fig:fan}, which lends support to approaches based on quantum criticality and fluctuations of the AFM/ORT phase.\cite{varma_fluctuations_2016, Lederer4905, fanfarillo_unconventional_2012} The fact that the superconducting dome precisely matches the ``shadow" of the critical fan on the zero temperature axis provides a surprisingly direct and intriguing connection between superconductivity and strange metal behavior. Previous studies have related superconductivity to various anomalous temperature dependencies in transport quantities,\cite{Abdel-Jawad_correlation_2007, cooper_anomalous_2009, Dorion-Leyraud_correlation_2009, jin_link_2011, mandal_anomalous_2019} but the use of a field-dependent characterization of strange metal behavior allows us to directly identify the limit of zero temperature as being relevant to the pairing interaction. Correlations between strange metal behavior and superconductivity are often taken as evidence that the pairing interaction also gives rise to strange metal behavior. However, on the overdoped side the strange metal behavior in the Hall effect of \basp\ gets stronger as temperature increases from zero, so the pairing interaction cannot be simply identified with the dynamics responsible for  strange metal corrections to the Hall effect. 
Rather, the pairing interaction seems to derive from just that part of the strange metal dynamics that is accessed via quantum fluctuations at zero temperature. This observation, which provides a new type of stringent test for theories of HTSC, would not have been possible without the ability to separate the low field enhancement of $R_H$, which is ostensibly a measure of  the role of correlations at different temperatures and compositions. 

The fact that this neat correlation appears in the Hall coefficient may not be an accident. The units of $R_H$ are inverse carrier density and a number of recent studies have indicated that the transition to a non-superconducting state in unconventional superconductors can be thought of as a gradual transference of electrons out of the condensate.\cite{bozovic_dependence_2016, paglione_quantum_2016, mahmood_locating_2018, moir_multi-band_2019} It is possible that the suppression of the strange metal correction to $R_H$ on the overdoped side is a reflection of the fact that fewer electrons are participating in the anomalous dynamics that are responsible for superconductivity. 

The observation that the strange metal contribution to $R_H$ simply adds to the band contribution is in tension with many approaches to understanding the strange metal state. For instance, it makes it unlikely that this behavior arises from a change in the effective carrier density with temperature, as that would influence $R_H$ through the low-field multiband formula (eq. 1 of the SI), leading to a different field dependence at different doping levels. 
Similarly, an anisotropic scattering rate should lead to different field dependences in $R_H$ as the mobilities and densities on the different fermi surface sheets change across the phase diagram. In contrast, we observe a constant field dependence in $R_H$ as a function of doping within a critical fan.
Many theoretical perspectives have been motivated by the observation that the cotangent of the Hall angle, $\rho_{xx}/R_H$, is approximately $\sim T^2$ in many HTSCs.\cite{chien_effect_1991, hwang_scaling_1994} However, according to our analysis the strange metal contribution to $R_H$ decouples from the band contribution in \basp, which casts doubt on any approach which treats the power law of the Hall angle as a fundamental quantity. On the other hand, some recent approaches to transport in strongly correlated systems do lead to linearly additive terms in $R_H$, consistent with our observations, and these may be relevant to high$-T_c$ superconductivity.\cite{auerbach_equilibrium_2018}

These observations of the field dependence $R_H$ significantly improve our picture of the relationship between HTSC and quantum criticality. They reveal the classic signature of quantum critical physics with surprising clarity, which is a welcome development since the mapping of the cuprate phase diagram onto the textbook quantum critical scenario has always been problematic.\cite{hwang_scaling_1994, ando_electronic_2004, cooper_anomalous_2009, hussey_contrasting_2018} On the other hand, our data appear to suggest a surprising upshot -  the influence of critical physics persists away from the critical point even at low temperatures, and understanding this will likely require a new theoretical approach.\cite{sondhi_continuous_1997} We emphasize that the present observations are model-independent, and thus provide a robust connection between metallic quantum criticality and high temperature superconductivity that should lead to a deeper understanding of both of these problems.

We would like to thank Dung-Hai Lee and Chandra Varma for fruitful discussions.
This work is supported by the Gordon and Betty Moore Foundation’s EPiQS Initiative, Grant GBMF9067. A portion work was performed at the National High Magnetic Field Laboratory, which is supported by National Science Foundation Cooperative Agreement No. DMR-1157490 and DMR-1644779 and the State of Florida. B.R., M-K.C. and R.D.M. acknowledge funding from the U.S. Department of Energy Office of Basic Energy Sciences Science at 100 T program.

\pagebreak

\end{document}